\documentclass[cits]{PoS}

\newcommand\url{}%
\newcommand\arcmin{\mbox{$^\prime$}}%
\newcommand\arcsec{\mbox{$^{\prime\prime}$}}%

\title{Living life on the edge - Wide-field VLBI at 90 cm! }

\ShortTitle{Living life on the edge - Wide-field VLBI at 90 cm! }

\author{\speaker{Emil Lenc}\\
        Swinburne University of Technology\\
        E-mail: \email{elenc@astro.swin.edu.au}}

\author{Mike Garrett, Olaf Wucknitz and James Anderson\\
        Joint Institute for VLBI in Europe\\
        E-mail: \email{garrett@jive.nl}, \email{wucknitz@jive.nl}, \email{anderson@jive.nl}}

\author{Steven Tingay\\
        Swinburne University of Technology\\
        E-mail: \email{stingay@astro.swin.edu.au}}

\abstract{We report on a recent 90 cm wide-field VLBI survey of two 3.1 deg$^{2}$ fields using the VLBA, Westerbork and Jodrell Bank telescopes. In-beam calibration was used to calibrate each field, the process was simplified by imaging the calibrators in DIFMAP and transferring the calibration solutions to AIPS using the newly developed DIFMAP task - \emph{cordump}. We detected and imaged 13 out of the 141 sources originally detected by the low resolution ($54\arcsec$) WENSS survey of the same two fields. The sources were detected at $7-12$ sigma levels above the image noise, had total flux densities ranging between 85$-$1640 mJy and were between $16\arcmin-58\arcmin$ from the phase centre of each field. This is the first systematic (and non-biased), deep, high resolution survey of the low frequency radio sky. These initial results suggest that new instruments such as LOFAR should detect many compact radio sources and that plans to extend these arrays to baselines of several thousand kilometres are warranted.}

\FullConference{8th European VLBI Network Symposium\\
		 September 26-29, 2006\\
		 Toru\'n, Poland}

\begin{document}

\section{Introduction}

Very little is known about the general properties of the 90 cm sky and even less is known at VLBI resolution. Previous snapshot surveys at these wavelengths have only targeted the brightest sources and were plagued by poor sensitivity and limited coherence times. As a result only a few tens of sources have been imaged with VLBI. Improved calibration and imaging techniques are required to learn more about the low frequency sky at VLBI resolutions. For example, Garrett et al. \cite{gar05} performed a deep VLBI survey at 20 cm of a $36\arcmin$ wide field by using a central bright source as an in-beam calibrator. The approach permitted the imaging of many potential target sources simultaneously and was ideal for survey work. We have set out to apply a similar technique at 90 cm by piggy-backing on a VLBI observation of the gravitational lens $0218+357$ and the nearby phase reference source J0226+3421 with the aim of surveying a 3 deg$^{2}$ field around each of the sources. The results provide important information with regards to what may be seen by future low-frequency instruments such as LOFAR, extended LOFAR and the SKA.

\section{Observation and Data Reduction}
A VLBI observation of the gravitational lens 0218$+$357 and the nearby phase calibrator J0226+3421 was made on 2005 November 11 using the Jodrell Bank, Westerbork and VLBA telescopes. The recorded data were correlated at JIVE in multiple passes to create a single-IF (4 MHz), single polarisation, wide-field data set with an integration time of 0.25 s and a multi-band, dual polarisation, narrow-field data-set with higher sensitivity. The wide-field data-set has a one sigma theoretical thermal noise level of $\sim$0.7 and $\sim$1.2 mJy/beam for the target and phase calibrator respectively.

Initial calibration was performed on the narrow-field data-set to take advantage of the increased sensitivity available with the additional bands and polarisations. Instrumental delays, which were assumed to be constant throughout the observation, were calibrated by fringe fitting on 3C84. A multi-band fringe fit was then performed on the phase calibrator. The resulting solutions were transferred to the wide-field data set using a ParselTongue\footnote{A Python scripting tool for AIPS available for download at \url{http://www.radionet-eu.org/rnwiki/ParselTongue}} script and the bandpass calibrated against observations of 3C84. Additional amplitude and phase calibrations were complicated by the complex structure of the phase calibrator J0226$+$3421. To simplify calibration, the phase calibrator data was averaged in frequency and exported to DIFMAP. Several iterations of modelling and self-calibration were performed in DIFMAP and a new DIFMAP task, \emph{cordump}\footnote{The \emph{cordump} patch is available for DIFMAP at \url{http://astronomy.swin.edu.au/~elenc/DifmapPatches/}}, was used to transfer the resulting phase and amplitude calibrations back to the unaveraged AIPS data-set. After application of these corrections the DIFMAP model of the source was subtracted from the AIPS u-v data-set to reduce side-lobe effects. The resulting data-set had a measured RMS noise approximately 1.8 times the theoretical thermal noise. The higher than anticipated noise is attributed to substantial levels of RFI observed on some baselines. With the phase calibrator solutions applied, the 0218+357 data set was also averaged in frequency and exported to DIFMAP to further refine the calibration in that field using the same process that was applied to the phase calibrator.

The data from both fields was kept in an unaveraged form to prevent smearing effects during imaging. Six sub-fields were defined for each field based on the distance from phase centre in $10\arcmin$ increments. The WENSS catalogue was used as a guide for potential targets in each sub-field. Dirty maps covering an area greater than that of the WENSS beam were created for each of the potential targets in each sub-field using the IMAGR task in AIPS. Heavier u-v tapering was applied to the outer sub-fields to reduce smearing effects but at the expense of reduced resolution and sensitivity. A 7$\sigma$ detection limit above the measured noise level was imposed to identify positive detections.

Phase corrections were observed to vary significantly between sub-fields as a result of ionospheric variations. A ParselTongue library has been written to interpolate and apply ionospheric total electron content (TEC) corrections to account for the spatially varying effect of the ionosphere across the field being imaged. However, the corrections did not result in any significant improvement in the resulting images. We suspect that the currently available TEC solutions may be too coarse both spatially and temporally to account for the ionospheric variations across the field.

\section{Results}
\begin{figure}
\includegraphics[width=\textwidth]{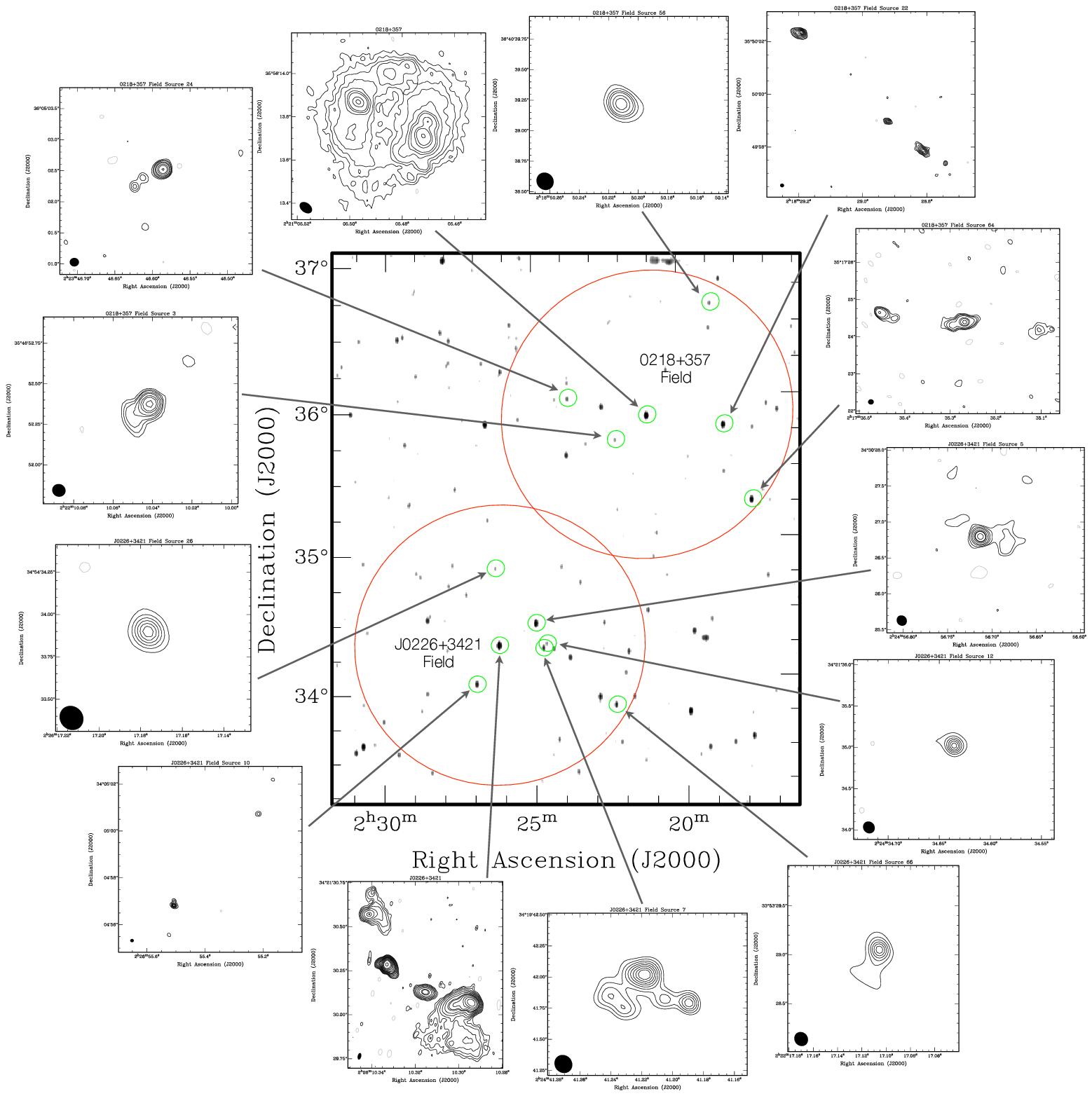}
\caption{Source detections using wide-field VLBI at 90 cm in 3.1 degree$^{2}$ fields around J0226$+$3421 and 0218$+$357.}
\label{fig1}
\end{figure}

The phase calibrator and target sources were imaged with a resolution of approximately 20 mas, both exhibited complex large scale structure. As shown in Figure \ref{fig1}, a total of 13 sources (including the central field sources) were detected from a total of 141 target sources with distances of up to a degree from the phase centre and at resolutions of between 80 and 140 mas. Six wide-field sources were detected out of 73 potential targets in the J0226$+$3421 field at $7-11\sigma$ levels, with total fluxes between 85$-$1640 mJy and with distances between $18\arcmin-56\arcmin$ from the phase centre. Five wide-field sources were detected out of 66 potential targets in the 0218$+$357 field at $7-12\sigma$ levels, with total fluxes between 87$-$1400 mJy and with distances between between $16\arcmin-58\arcmin$ from the phase centre. Almost all of the detected sources were extended suggesting that those sources that were not detected were resolved out by the significantly smaller beam of this observation compared to that of WENSS ($\sim1\arcmin$). Nonetheless, our observations have shown that there are many sources that could be studied at longer wavelengths with instruments such as LOFAR, extended LOFAR and SKA.

\section{Conclusion}
Approximately 10\% of WENSS survey sources within two 3 deg$^{2}$ fields have been detected and imaged using wide-field VLBI imaging techniques at 90 cm. The results provide an indication of what may be observed by next generation low-frequency instruments such as LOFAR and SKA, they also suggest that plans to extend these arrays to baselines of several thousand kilometres are warranted. It would be informative to apply the wide-field techniques described here to a larger sample of fields. Such a study could better constrain the frequency and nature of sources that may be targeted by instruments such as LOFAR.

\end{document}